# Post-spinel transformations and equation of state in ZnGa$_2$O$_4$: Determination at high-pressure by in situ x-ray diffraction

D. Errandonea[1,*], Ravhi S. Kumar[2], F. J. Manjón[3], V. V. Ursaki[4], and E. V. Rusu[4]

[1]MALTA Consolider Team, Departamento de Física Aplicada-ICMUV, Fundación General de la Universitad de Valencia, Edificio de Investigación, c/Dr. Moliner 50, 46100 Burjassot (Valencia), Spain

[2]High Pressure Science and Engineering Center, Department of Physics and Astronomy, University of Nevada Las Vegas, 4505 Maryland Parkway, Las Vegas, Nevada 89154-4002, USA

[3]MALTA Consolider Team, Departamento de Física Aplicada-IDF, Universitat Politècnica de València, Cno. de Vera s/n, 46022 València, Spain

[4]Institute of Applied Physics, Academy of Sciences of Moldova, 2028 Chisinau, Moldova

**Abstract:** Room temperature angle-dispersive x-ray diffraction measurements on spinel ZnGa$_2$O$_4$ up to 56 GPa show evidence of two structural phase transformations. At 31.2 GPa, ZnGa$_2$O$_4$ undergoes a transition from the cubic spinel structure to a tetragonal spinel structure similar to that of ZnMn$_2$O$_4$. At 55 GPa, a second transition to the orthorhombic marokite structure (CaMn$_2$O$_4$-type) takes place. The equation of state of cubic spinel ZnGa$_2$O$_4$ is determined: $V_0 = 580.1(9)$ Å$^3$, $B_0 = 233(8)$ GPa, $B_0' = 8.3(4)$, and $B_0'' = -0.1145$ GPa$^{-1}$ (implied value); showing that ZnGa$_2$O$_4$ is one of the less compressible spinels studied to date. For the tetragonal structure an equation of state is also determined: $V_0 = 257.8(9)$ Å$^3$, $B_0 = 257(11)$ GPa, $B_0' = 7.5(6)$, and $B_0'' = -0.0764$ GPa$^{-1}$ (implied value). The reported structural sequence coincides with that found in NiMn$_2$O$_4$ and MgMn$_2$O$_4$.

* Corresponding author, Email: daniel.errandonea@uv.es, Fax: (34) 96 3543146, Tel.: (34) 96 354 4475

## I. Introduction

Cubic oxide spinel $AM_2O_4$ compounds (A: bivalent cation and M: trivalent cation) occur in many geological settings of the Earth's crust and mantle, as well as in lunar rocks and meteorites. The study of their high-pressure structural properties is important for improving the understanding of the constituents of the Earth. High-pressure studies have been performed in $MgM_2O_4$ spinels (e.g. $MgAl_2O_4$) revealing that upon compression they may adopt orthorhombic $CaFe_2O_4$-, $CaMn_2O_4$-, or $CaTi_2O_4$-type structures [1]. However, the structure and properties of post-spinel phases is presently still under debate. On top of $MgM_2O_4$ spinels, the high-pressure properties of $ZnM_2O_4$ cubic spinels (e.g. $ZnAl_2O_4$) have been studied too. Among them, $ZnAl_2O_4$ [2] and $ZnFe_2O_4$ [3] have been experimentally investigated. The first one remains stable up to 43 GPa in the cubic spinel structure but the second one transforms to either a $CaFe_2O_4$- or a $CaTi_2O_4$-type structure beyond 24 GPa. In addition to these facts, in other compounds like $AMn_2O_4$ spinels, cubic-to-tetragonal transitions have been reported to occur at pressures as low as 12 GPa [4]. In contrast with the materials above mentioned, the high-pressure structural stability of $AGa_2O_4$ spinels has not been studied yet. In order to shed more light on the understanding of the high-pressure properties of $AM_2O_4$ cubic spinels, we report a study of the high-pressure structural properties of zinc gallate ($ZnGa_2O_4$) up to 56 GPa. The present research work contributes to achieve a fuller understanding of how cation replacement affects the high-pressure behavior of oxide spinels.

## II. Experimental details

$ZnGa_2O_4$ powders were synthesized by a solid state reaction at high temperature by mixing appropriate quantities of ZnO and $Ga_2O_3$ precursors and firing at 1100ºC for 24 h [5]. Chemical and structural analyses have shown the stoichiometric composition of ZnGa$_2$O$_4$ and the presence of traces of impurities (β-Ga$_2$O$_3$, less than 1% by volume, and ZnO, less than 0.2% by volume). Angle-dispersive x-ray diffraction (ADXRD) experiments were carried out at room temperature (RT) at high pressure up to 56 GPa at Sector 16-IDB of the HPCAT - Advanced Photon Source (APS) - using a diamond-anvil cell (DAC) and an incident monochromatic wavelength of 0.36816(5) Å. Samples were loaded in a 100 μm hole of a rhenium gasket in a Mao-Bell-type DAC with diamond-culet sizes of 300 μm. Ruby grains were loaded with the sample for pressure determination [6] and silicone oil was used as pressure-transmitting medium [7, 8]. The monochromatic x-ray beam was focused down to 20 × 20 μm$^2$ using Kickpatrick-Baez mirrors. The images were collected using a MAR345 image plate located at 350 mm from the sample. They were integrated and corrected for distortions using FIT2D. The structural analysis was performed using POWDERCELL.

**III. Results and discussion**

At ambient pressure the wide band-gap semiconductor ZnGa$_2$O$_4$ has a cubic spinel structure (space group: $Fd\bar{3}m$). The spinel AM$_2$O$_4$ structure has two symmetrically distinct cation sites, with two octahedrally-coordinated M cations for each tetrahedrally-coordinated A cation. A cations ocuppy the Wyckoff position 8a at (1/8, 1/8, 1/8), M cations ocuppy the Wyckoff position 16c at (1/2, 1/2, 1/2), and O atoms are located in the Wyckoff position 32e at ($u$, $u$, $u$), where $u$ is approximately 1/4. The cubic spinel structure has only two structural variables, the unit-cell parameter $a$ and the oxygen position parameter $u$. Inversion (A and M atoms mutually substitute for one another) is a typical phenomenon in cubic spinels. "Normal" spinels have all A cations in the tetrahedral site and all M cations in the octahedral site, while "inverse" spinels are disordered on the octahedral site [M(AM)O$_4$ spinel]. Maximum distortion occurs on both sites for intermediate forms. ZnGa$_2$O$_4$ has been reported to be a "normal" spinel with $a = 8.3358(5)$ Å and $u = 0.2608(6)$ [9]. At ambient pressure (and outside the DAC), the x-ray diffraction pattern of our samples indicated that they have a "normal" spinel structure with $a = 8.335(8)$ Å and $u = 0.2600(6)$. This structure will be named along the paper as c-ZnGa$_2$O$_4$. A slight degree of inversion could be present in c-ZnGa$_2$O$_4$ but we cannot determine it precisely because Zn and Ga atoms have nearly equal x-ray scattering factors. However, the effects of this slight inversion are smaller than uncertainties introduced in high-pressure measurements. So we will neglect cation inversion in the analysis of our high-pressure data. The small amount of impurities present in our samples was detected in high-resolution diffraction patterns measured outside the DAC. Inside the DAC, only weak peaks from β-Ga$_2$O$_3$ impurities were detected at low pressures. However, they do not preclude the clear identification of the ZnGa$_2$O$_4$ peaks. The most intense peaks of β-Ga$_2$O$_3$ are indicated with asterisks in the lower trace of Fig. 1.

Fig. 1 shows representative diffraction patterns measured up to 56 GPa. Up to 31.2 GPa the patterns can be indexed with the cubic spinel structure. The diffraction peaks broaden upon compression but the patterns are good enough to allow a Rietveld refinement up to 25.7 GPa. From the refinements we obtained the evolution of the unit-cell and oxygen position parameters as a function of pressure. Fig. 2 shows the pressure dependence of the volume (only the squares correspond to c-ZnGa$_2$O$_4$). A fit to the data reported up to 25.7 GPa with a third-order Birch-Murnaghan equation of state (EOS) [9] gives: $V_0 = 580.1(9)$ Å$^3$, $B_0 = 233(8)$ GPa, and $B_0' = 8.3(4)$, where $V_0$, $B_0$, and $B_0'$ are the zero-pressure volume, bulk modulus, and its pressure derivative, respectively. According with these parameters the implied value of the second derivative of the bulk modulus versus pressure at ambient conditions is $B_0'' = -0.1145$ GPa$^{-1}$. The EOS fit is shown as a solid line in Fig. 2 and the agreement with the experiments is found to be good. According to this result $ZnGa_2O_4$ is the less compressible oxide spinel among those studies up to now (see Table I) [2, 3, 11 – 16]. On the other hand, the estimated $B_0$ is in good agreement with the theoretical predictions made using the self-consistent tight-binding linearized muffin-tin orbital (LMTO) method [17, 18], the *ab initio* perturbed ion (aiPI) model [19, 20], or the local-density approximation (LDA) [21, 22]. All these theoretical methods give values for $B_0$ ranging from 207 to 243 GPa, which include our experimental value (see Table II). In contrast with other theoretical methods, the generalized gradient approximation (GGA) gives an underestimated value for $B_0$ [22, 23] as can be seen in Table II. It should be said here, that a partial inversion between Zn and Ga sites could cause a decrease of the compressibility of the $ZnO_4$ tetrahedra. A decrease of this compressibility would necessary lead to an enhancement of $B_0$ if we compare it with that of the normal spinel [19]. Evidence of such pressure-induced cation inversion has been found in other oxide spinels upon compression; e.g. $NiAl_2O_4$ [24]. However, as we discussed above, inversion cannot be detected in $ZnGa_2O_4$ due to the similar x-ray scattering factors of Zn and Ga. Therefore, the value of 233(8) GPa should be considered as an upper limit for $B_0$ in c-$ZnGa_2O_4$.

From our data we also deduce a negative pressure coefficient of the parameter *u* under compression. This tendency indicates that spinel tries to reach the ideal structure ($u = 0.25$) when pressurized. The pressure dependence of *u* is shown in the inset on Fig. 2, together with a linear fit [$u = 0.2599(6) - 0.00009(5)P$, where *P* is in GPa]. It is also compared with theoretical calculations [23]. Both, calculations and experiments, give a decrease of *u* under compression. A similar behavior was observed in other cubic spinels like $ZnAl_2O_4$ [2]. The calculations predict a similar behavior for $ZnAl_2O_4$ and $ZnIn_2O_4$ [23]. The change of *u* with pressure makes the polyhedral compressibility of $ZnO_4$ tetrahedra to be larger than that of the $GaO_6$ octahedra. To understand this fact we should remember that the Zn-O and Ga-O bond distances are given by: $R_{Zn-O} = \sqrt{3}\, u - 0.125\, a$ and $R_{Ga-O} = \sqrt{3u^2 - 2u + 0.375}\, a$. At $u = 0.2625$ $R_{Zn-O} = R_{Ga-O}$ and at $u = 0.25$ the oxygen atoms form a perfect face centered cubic sublattice, being the following relations obtained: $\partial R_{Zn-O} / \partial u = \sqrt{3} a$ and $\partial R_{Ga-O} / \partial u = -a$. Thus when $u$ increases, the tetrahedral bond distances increase while the octahedral bond lengths decrease, and the tetrahedral bond length change faster than the octahedral one. Therefore, upon compression we have two competing effects. On one side, both bond distances tend to decrease because of the reduction of the unit-cell parameter. On the other hand, the reduction of $u$ contributes to the decrease of the Zn-O distance but partially compensate the decrease of the Ga-O distance (because $\partial R_{Ga-O} / \partial u = -a$). Thus the Zn-O bonds are expected to be more compressible than the Ga-O bonds, which is in excellent agreement with our experiments. Fig. 3 shows the pressure evolution of the bond distances (Zn-O = 1.949 Å and Ga-O = 2.004 Å at ambient pressure according to our data). It can be seen there, that the Zn-O bond is more compressible than the Ga-O bond. The pressure coefficients are -0.0017(2) GPa$^{-1}$ and –0.0007(2) GPa$^{-1}$, respectively. A similar differential bond compressibility was previously reported in other spinels like MgAl$_2$O$_4$ [25] and ZnAl$_2$O$_4$ [2]. Furthermore, the difference found in bond compressibilities gives an octahedral compressibility of 265(9) GPa and a tetrahedral compressibility of 155(9) GPa. This difference in polyhedral compressibility is in good agreement with the predictions of Recio *et al.* [19]. Indeed, our results confirm that the bulk compressibility of ZnGa$_2$O$_4$ can be expressed in terms of the polyhedral compressibilities [19].

Changes in the diffraction patterns are found at 31.2 GPa. Peaks become much broader and many of them split, as indicated by arrows in Fig. 1. Apparently a phase transition takes place in $ZnGa_2O_4$ at 31.2 GPa. The splitting of the peaks becomes more notorious at higher pressures as can be seen in the spectra recorded at 34.8 and 48.7 GPa. The changes of the diffraction patterns cannot be explained by any of the known high-pressure post-spinel phases ($CaFe_2O_4$-, $CaMn_2O_4$-, and $CaTi_2O_4$-type). A possible explanation to the experimental results is given by the consideration of the tetragonal spinel structure, the structure of $ZnMn_2O_4$ and $MgMn_2O_4$ [12, 26]. This phase belongs to space group $I4_1/amd$, a translationengleiche subgroup of $Fd\bar{3}m$. Thus, we propose that the tetragonal high-pressure phase is formed by a tetragonal distortion of the cubic spinel (if in the tetragonal structure $c/a = \sqrt{2}$ is satisfied, it is reduced to the cubic spinel). At 31.2 GPa we obtained for the tetragonal phase $a = 5.743(6)$ Å and $c = 8.032(8)$ Å, $c/a = 1.398(2)$. The diffraction patterns measured up to 48.7 GPa can be also indexed with the tetragonal spinel structure (t-$ZnGa_2O_4$). On indexing we obtained the pressure evolution of $a$, $c$, and the volume. The results obtained for $a$ and $c$ are shown in Fig. 4 (full circles upstroke and open circles downstroke) and those obtained for the volume are shown in Fig. 2 (circles). Since the cubic phase has $Z = 8$ and the tetragonal phase has $Z = 4$, to compare them we have considered V' = 2V for t-$ZnGa_2O_4$. We found that the $c$ axis is more compressible than the $a$ axis. Thus, upon compression the tetrahedral distortion of the structure increases (and consequently the splitting of the diffraction peaks), as can be seen in the inset of Fig. 4 where we show $c/a$ versus pressure.

An interesting scenario is that the EOS of c-$ZnGa_2O_4$ describes well the compressibility of t-$ZnGa_2O_4$. To illustrate it, the EOS of c-$ZnGa_2O_4$ is extrapolated as a dotted line in Fig. 2. A fit to the data reported for the tetragonal phase up to 48.7 GPa (obtained both, on compression and decompression) with a third-order Birch-Murnaghan equation of state (EOS) [9] gives: $V_0 = 257.8(9)$ Å$^3$, $B_0 = 257(11)$ GPa, and $B_0´= 7.5(6)$, being the implied $B_0´´= -0.0764$ GPa$^{-1}$. The agreement of this EOS with the experimental data is similar to that found in c-ZnGa$_2$O$_4$. The EOS is not represented in Fig. 2 for the sake of clarity. In addition, apparently within the accuracy of the experiments, there is no volume change at the transition. It is important to note, that the cubic-tetragonal transition has been documented previously in NiMn$_2$O$_4$ at 12 GPa [15]. These facts, and the group-subgroup relationship existent between the cubic and tetragonal phases, point toward the occurrence of a second-order pressure-induced transition in ZnGa$_2$O$_4$ [21]. Regarding cation coordination, t-ZnGa$_2$O$_4$ also consists of ZnO$_4$ tetrahedra and GaO$_6$ octahedra, but in contrast with c-ZnGa$_2$O$_4$, the octahedra are distorted showing four long distances and two shorter distances. At 31.2 GPa the Zn-O distance is 1.843 Å and the Ga-O distances are 1.966 Å and 1.945 Å.

Further compression gives rise to a second transition at 55.4 GPa. At this pressure there are important changes observed in the diffraction patterns. The number of diffraction peaks increases considerably and the patterns resemble those of the orthorhombic post-spinel structures for MgAl$_2$O$_4$ and homologous compounds [27]. We tried to index the pattern collected at 55.4 GPa with the CaFe$_2$O$_4$-type (space group: *Pnma*), CaMn$_2$O$_4$-type (space group: *Pbcm*), and CaTi$_2$O$_4$-type (space group: *Cmcm*) structures. The best agreement between the observed and calculated patterns has been achieved when considering the orthorhombic CaMn$_2$O$_4$-type structure (marokite) [28]. At 55.4 GPa, we obtained *a* = 2.93(1) Å, *b* = 9.13(3) Å, and *c* = 8.93(3) Å for orthorhombic ZnGa$_2$O$_4$ (o-ZnGa$_2$O$_4$). The same structure was found to be the high-pressure phase of the tetragonal spinel MgMn$_2$O$_4$ [26]. The transition from t-ZnGa$_2$O$_4$ to o-ZnGa$_2$O$_4$ involves an estimated volume collapse of around 7 % at the transition; i.e. it is a first-order transition. The compressibility of o-ZnGa$_2$O$_4$ cannot be extracted from the present experiments, but given its density and cation coordination increase, it is expected to be less compressible than the low-pressure phases.

As regards the cation coordination, an increase of the Zn coordination in $ZnGa_2O_4$ occurs during the transition from t-$ZnGa_2O_4$ to o-$ZnGa_2O_4$. The orthorhombic structure is made up of $GaO_6$ distorted octahedra and $ZnO_8$ zinc centered polyhedra [28]. The abrupt increase in zinc coordination from 4 to 8 at 55 GPa is rather remarkable if we consider that the high-pressure rocksalt phase of ZnO, obtained above 8-9 GPa and where Zn is octahedrally-coordinated, is stable up to at least 200 GPa [29]. It is interesting to note that c-$ZnGa_2O_4$ and t-$ZnGa_2O_4$ show Zn-O bond distances that are smaller than that found for the average distance in wurtzite-type ZnO at the wurtzite-to-rocksalt phase transition (1.951 Å). This means that the cubic and tetragonal spinel structures extend the fourfold coordination to very high pressures (31 GPa). The same phenomena occurs in $ZnAl_2O_4$ where the cubic spinel structure is stable up to 43 GPa [2]. On the other hand, the average value of the Zn-O distances in o-$ZnGa_2O_4$ is 2.21 Å (the different distances range from 2.085 Å to 2.425 Å), which is larger than the highest and the smallest Zn-O distances found in rocksalt-type ZnO (2.141Å at ambient pressure and 1.806Å at 200 GPa). This means that the orthorhombic marokite structure allows the stabilization of eightfold-coordinated Zn at rather low pressures. In contrast with the behavior of the Zn-O distances. The Ga-O distances in the different phases of $ZnGa_2O_4$ correlate well with those of the different high-pressure phases of $Ga_2O_3$ [30].

Upon decompression $ZnGa_2O_4$ reverts from o-$ZnGa_2O_4$ to t-$ZnGa_2O_4$ at 35 GPa, but it does not revert to c-$ZnGa_2O_4$ at ambient pressure. The same scenario has been observed in $NiMn_2O_4$ [15]. At ambient conditions the parameters of the metastable tetragonal phase are $a$ = 5.914(6) Å and $c$ = 8.221(8) Å. The volume (V' = 2V) agrees within accuracy of experiments with that of the cubic phase. An interesting fact is that on pressure release the *c/a* ratio does not follow the inverse behavior observed on pressure increase. Apparently when releasing pressure *c/a* is much less affected by pressure, reaching a value close to 1.39 at ambient pressure, while on compression we started with a value of $c/a = 1.398(2)$ at 31.2 GPa when t-$ZnGa_2O_4$ was first detected. The reason of this different behavior is not understood yet, but it can be related either with non-hydrostatic effects [7, 31], which may preclude the recovering of the tetragonal distortion induced by pressure, or with a pressure-induced inversion in the cation sites after the second phase transition, which could be related to the fact that some Zn cations remain octahedrally coordinated on releasing pressure. Note that a smaller c/a value was found in partially inverted tetragonal spinel $MgMn_2O_4$ at ambient pressure as compared to the normal spinel [26]. These effects could be also the reason that avoids the recovering of c-$ZnGa_2O_4$, which should be expected if the cubic-tetragonal transition has a second-order character, as tentatively proposed here.

To conclude we would like to mention that the reported pressure for cubic-tetragonal transition agrees with the predictions made based on the ratio between the average ionic radius of Zn and Ga and the ionic radius of O [32, 33]. This empirical criterion predicts a transition pressure of 33(3) GPa for $ZnGa_2O_4$, very close to our experimental value. We would also like to add that the mechanisms of this transition and the tetragonal-orthorhombic transition are still an open question. Further high-pressure studies on $ZnGa_2O_4$ using Raman spectroscopy, *ab initio* calculations, optical spectroscopy, and dielectric measurements are in progress to clarify this behavior and to understand the physical properties of the low- and high-pressure phases of $ZnGa_2O_4$. The combination of these techniques has proven to be an efficient tool to study the high-pressure properties of different minerals and semiconductors [34 – 38].

## IV. Summary

In summary, we performed RT ADXRD measurements on cubic spinel $ZnGa_2O_4$ up to 56 GPa. We found the occurrence of two post-spinel phase transitions at 31.2 GPa and 55.4 GPa respectively, and determined the EOS for c-$ZnGa_2O_4$: $V_0$ = 580.1(9) $Å^3$, $B_0$ = 233(8) GPa, and $B_0´$ = 8.3(4); implied $B_0''$ = -0.1145 $GPa^{-1}$. For this phase we found a differential polyhedral compressibility for $GaO_6$ octahedra and $ZnO_4$ tetrahedra as predicted in previous calculations [19]. This fact appears to be related with the change with pressure of the oxygen position parameter $u$. Regarding the high-pressure phases we propose a tetragonal spinel structure, similar to that of $MgMn_2O_4$, and an orthorhombic $CaMn_2O_4$-type structure. This structural sequence is the same as that in the results reported for $NiMn_2O_4$ [15] and $MgMn_2O_4$ [26]. *Ab initio* total-energy calculations are also consistent with this sequence [21]. Finally, for t-$ZnGa_2O_4$ we found that it has a similar compressibility than c-$ZnGa_2O_4$, being the EOS parameters: $V_0$ = 257.8(9) $Å^3$, $B_0$ = 257(11) GPa, and $B_0´$ = 7.5(6); implied $B_0''$ = -0.0764 $GPa^{-1}$.

**Acknowledgments:** Research financed by the Spanish MICINN and the Generalitat Valenciana (Grants: MAT2007-65990-C03-01, MAT2006-02279, CSD-2007-00045, GV06-151, and GV2008-112). The U.S. DOE, Office of Science, and Office of Basic Energy Sciences support the use of the APS (Contract: W-31-109-Eng-38). DOEBES, DOE-NNSA, NSF, DOD-TACOM, and the Keck Foundation support HPCAT. Work at UNLV is financed by DOE (Award: DEFG36-05GO08502). DOENNSA supports the UNLV HiPSEC (Agreement: DE-FC52-06NA26274). F.J.M. thanks the support from "Vicerrectorado de Innovación y Desarrollo - UPV" (Grant: UPV2008-0020).

**Table I:** Comparison of experimentally-determined bulk moduli of different oxide spinels. $ZnGa_2O_4$ is apparently the less compressible compound among them. The pressure derivative of the bulk modulus is also included for completeness.

| Compound | $B_0$ (GPa) | $B_0'$ | Reference | Compound | $B_0$ (GPa) | $B_0'$ | Reference |
|---|---|---|---|---|---|---|---|
| c-$ZnGa_2O_4$ | 233(8) | 8.3(4) | This work | $MgAl_2O_4$ | 194 - 212 | 4 – 5.6 | [13, 14] |
| t-$ZnGa_2O_4$ | 257(11) | 7.5(6) | This work | $MgFe_2O_4$ | 181 | 6.3 | [11] |
| $ZnAl_2O_4$ | 201 - 211 | 4.8 – 7.6 | [2] | $Mg_2SiO_4$ | 184 | 4 | [14] |
| $ZnCr_2O_4$ | 183 | 7.9 | [11] | $Ni_2SiO_4$ | 220 | 4 | [13] |
| $ZnFe_2O_4$ | 166 | 9.3 | [3] | $NiMn_2O_4$ | 206 | 4 | [15] |
| $ZnMn_2O_4$ | 183 | 4 | [12] | $CuMn_2O_4$ | 198 | 4 | [16] |

**Table II:** Experimentally and calculated bulk modulus of c-$ZnGa_2O_4$.

| Method | $B_0$ (GPa) | Reference | Method | $B_0$ (GPa) | Reference |
|---|---|---|---|---|---|
| ADXRD | 233(8) | This work | GGA | 156 | [23] |
| aiPI | 207.5 | [19, 20] | LDA | 220 | [21] |
| LMTO | 237 | [11] | GGA | 147 | [22] |
| LMTO | 243 | [18] | LDA | 217 | [22] |

**Figure captions**

**Figure 1:** X-ray diffraction patterns of $ZnGa_2O_4$ at selected pressures (indicated). In the lower trace asterisks show the peaks associated to β-$Ga_2O_3$. Arrows denote the peak splitting indicative of the onset of the cubic-to-tetragonal transition. The upper trace shows a pattern collected on pressure release to illustrate the non-reversibility of this transition.

**Figure 2:** Pressure evolution of the volume of $ZnGa_2O_4$. Squares: c-$ZnGa_2O_4$. Circles: t-$ZnGa_2O_4$ (solid: upstroke, empty: downstroke). The solid line represents the EOS of c-$ZnGa_2O_4$. The dotted line is its extrapolation from 25.7 GPa to 50 GPa. The inset gives the pressure evolution of the oxygen position parameter $u$. Symbols: present data. Solid line: linear fit to experiments. Dashed line: theoretical calculations [23].

**Figure 3:** Pressure dependence of the bond distances on c-$ZnGa_2O_4$. Circles: present data. Lines: linear fit to the data.

**Figure 4:** Pressure evolution for the unit-cell parameters of the cubic ($a_C$: squares) and the tetragonal ($a_T$, $c_T$: circles) phases. Full (empty) symbols were taken on compression (decompression). The inset shows the pressure evolution of the $c/a$ ratio. Lines are just a guide to the eye.

**Figure 1**

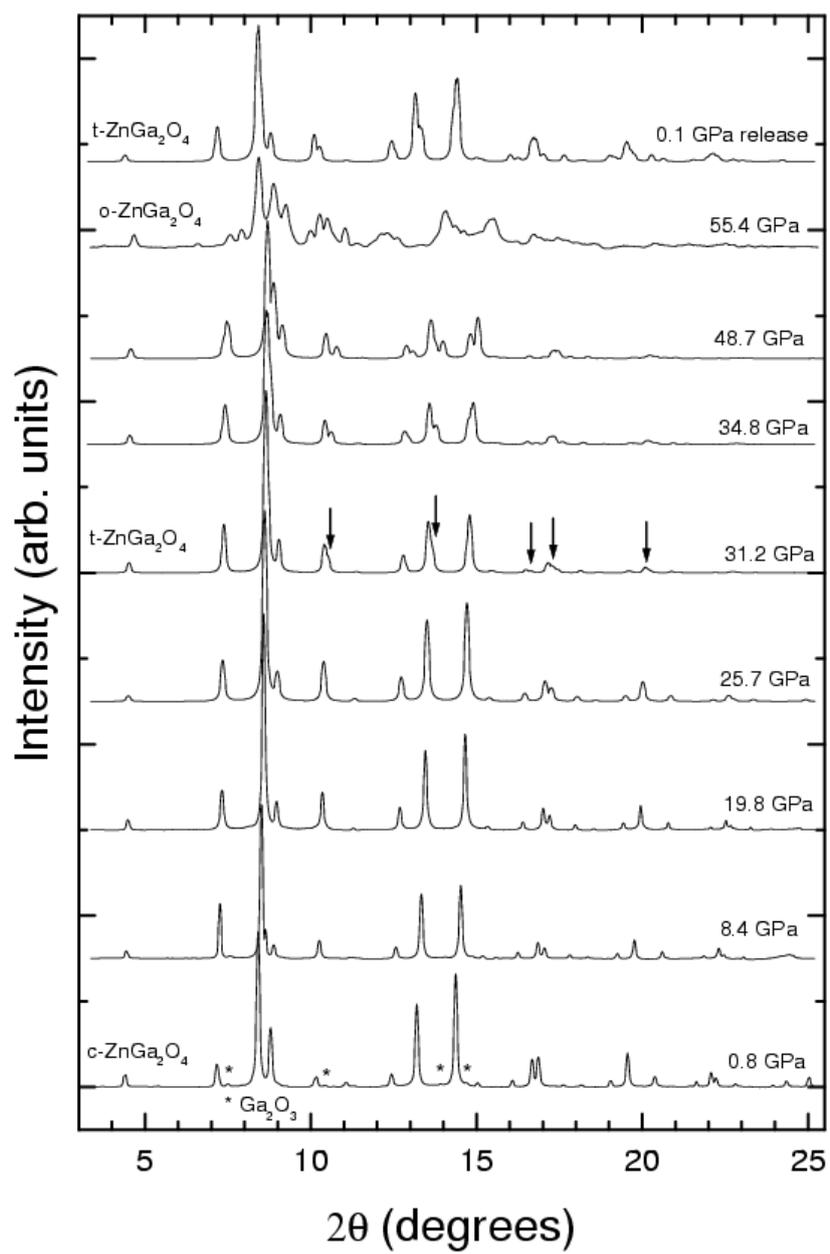

**Figure 2**

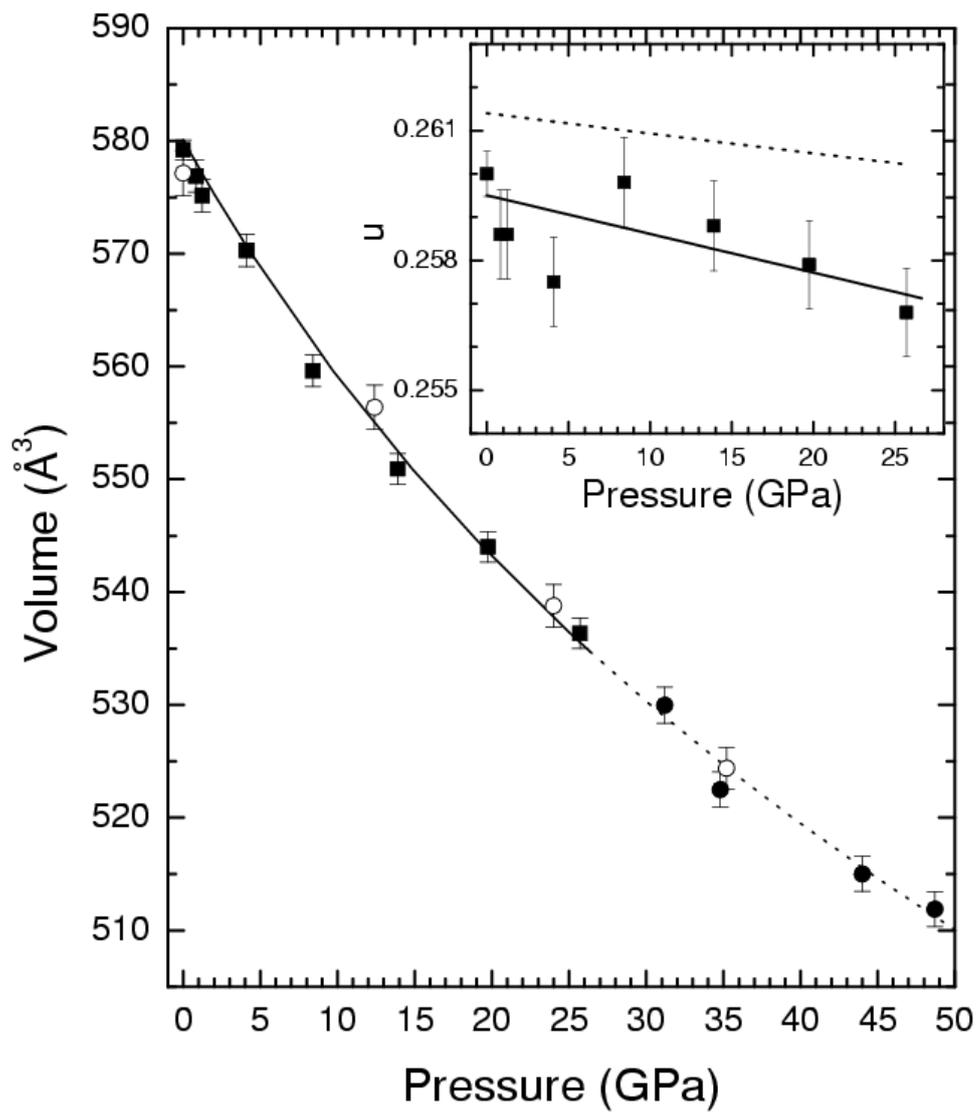

**Figure 3**

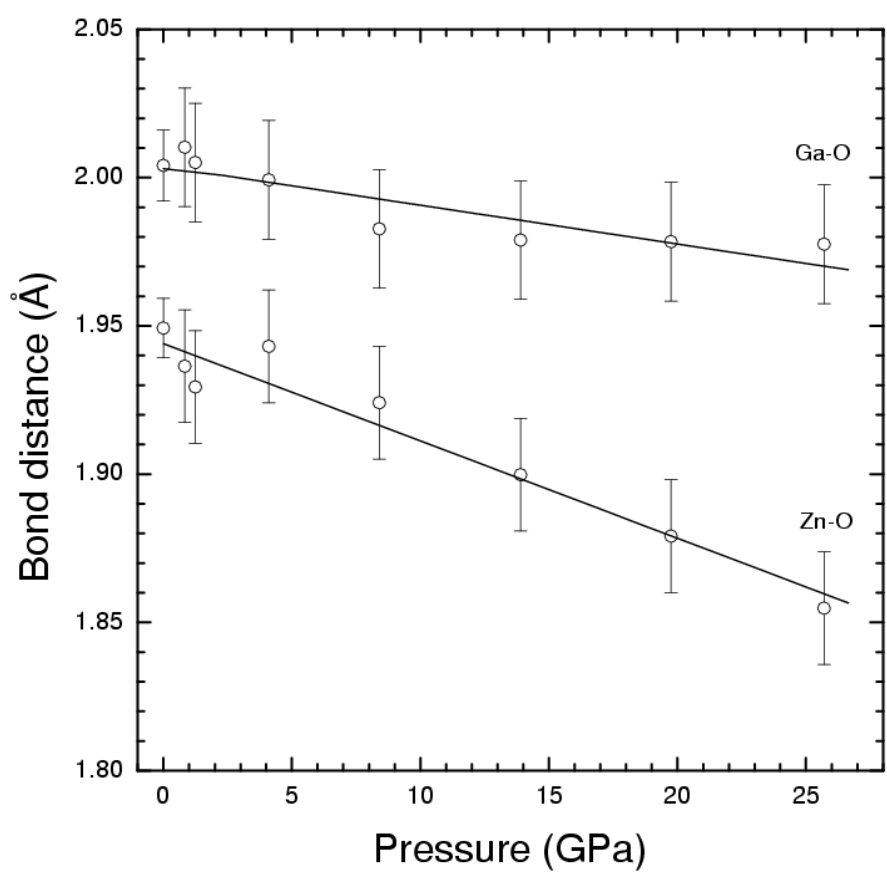

**Figure 4**

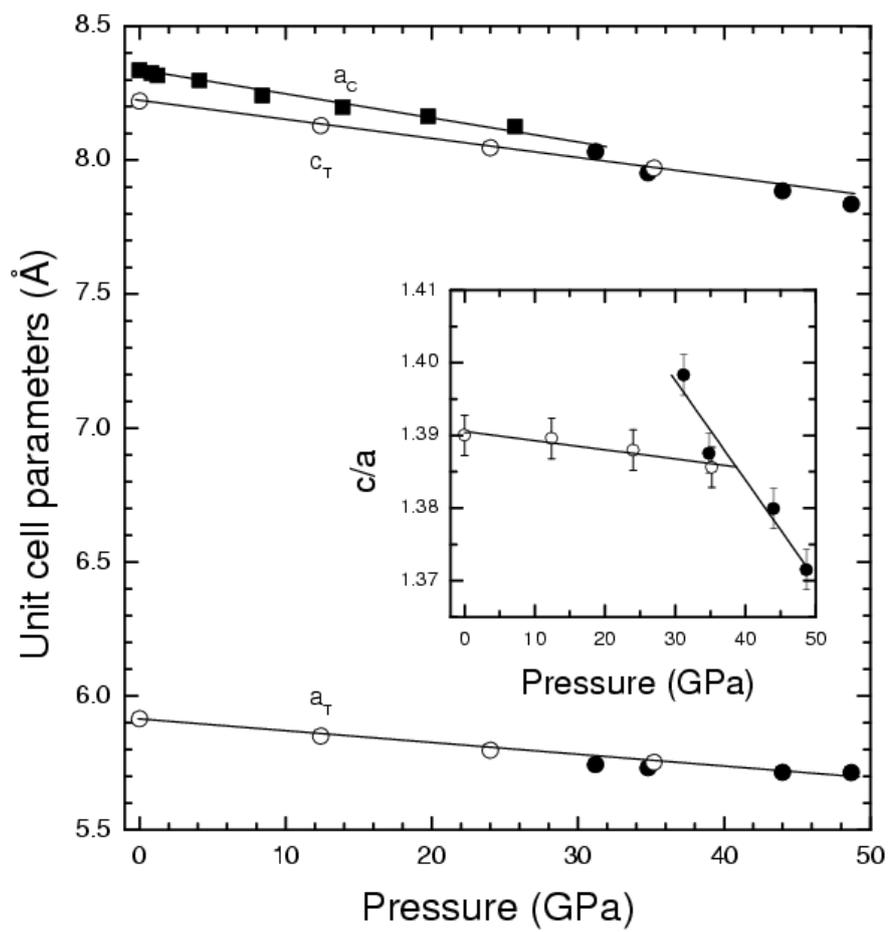